\documentclass[epj]{svjour}

\usepackage{latexsym}
\usepackage{graphics}
\usepackage{hyperref}
\usepackage{indentfirst}
\usepackage{bm}
\usepackage{cite} 
 \hypersetup{colorlinks=true, urlcolor=blue, 
linkcolor=blue, citecolor=blue} 
\usepackage{lineno,hyperref}
\usepackage{graphicx}
\usepackage{amsmath}
\usepackage{amssymb}
\usepackage{mathrsfs}
\usepackage{tikz}
\usepackage{filecontents}
\begin{document}
\title{The specific heat and magnetic properties of a spin-1/2 ladder including butterfly-shaped unit blocks}
\author{Hamid Arian Zad 
\thanks{\emph{}
e-mail: arianzad.hamid@mshdiau.ac.ir}
}
% etc
%\thanks{\emph{}
%e-mail: arianzad.hamid@mshdiau.ac.ir}

\institute{Young Researchers and Elite Club, Mashhad Branch, Islamic Azad University, Mashhad, Iran}
%\and the second here}
\abstract{
The specific heat, structural characterization, and magnetic property studies of a new spin ladder with the geometry of butterfly-shaped configuration are reported.
The model introduced here is an infinite spin ladder-type including spin-1/2 particle for which unit blocks consist of 
two  butterflies connected together through their bodies (Body-Body bridges).  Localized spins on the wings of butterflies have XXZ Heisenberg interaction with two extra spin-1/2 particles assumed in the center of each cage (unit block), while they have pure Ising-type interaction with those spins that are localized on the bodies. Hence, there are six interstitial spins and four nodal spins (Body-Body interaction) per block. To obtain the partition function of this model, we use the transfer matrix approach, then we examine the magnetization process, as well as, the specific heat of the model. Interestingly, we see a wide magnetization plateau at $\frac{5}{6}$ of the saturation magnetization that is strongly dependent on the magnetic field and anisotropy variations. Moreover, some unexpected phenomena are observed in the low-temperature limit, such as anomalous triple-peak in the specific heat function which gradually turns to a double-peak upon increasing the magnetic field and/or anisotropic Heisenberg coupling, due to the ferromagnetic phase predomination. }
%
%\PACS{
      %{}{}   %\and
   %   {PACS-key}{discribing text of that key}
   %  } % end of PACS codes
 %end of abstract
%
\maketitle
\section{Introduction}
Obtaining a profound perception about interacting quantum many-body systems like low-dimensional magnetic materials with competing interactions or geometrical frustration have become an intriguing research object in a number
of subjects such as condensed matter physics, material science and inorganic chemistry. In these particular areas
many investigations concerned about quantum ferrimagnetic chains  have been carried out, due to that they present a relevant combination of ferromagnetic and antiferromagnetic phases. As a result of zero and finite temperature phase transitions, these materials present various ground states and thermal properties \cite{Kitaev,Gu,Dillenschneider,Ivanov,Werlang,Sachdev,Rojas2,RojasM,Strecka}. Spin ladders can be count as attractive models among these systems. The latter consist of square-shaped topological units along the ladder \cite{Strecka,Koga,Okamoto,Muller,Vuletic,Notbohm,Buttner,Blundell,Bacq,Arian1,Feng2007}. 

During the past two decades it has become possible to synthesize a large variety of compounds such as $A_3Cu_3(PO_4)_4$ with $A=Ca, Sr$ \cite{Drillon1}, $Cu_3Cl_6(H_2O)_2 \cdot 2H_8C_4SO_2$ \cite{Okamoto1999,Okamoto2003}, the ferromagnetic diamond chains in polymeric coordination  compound $Cu_3(TeO_3)_2Br_2$ \cite{Uematsu} and the natural mineral azurite ($Cu_3(CO_3)_2(OH)_2$) \cite{Kikuchi1,Kikuchi2}, which can be properly introduced in terms of Heisenberg spin models. Recently,  A. Baniodeh {\it et al.} verified experimentally the ground state as well as low-temperature thermodynamic properties of material $\big[Fe_{10}Gd_{10}(Me-tea)_{10}(Me-teaH)_{10}(NO_3)_{10}\big]\cdot 20MeCN$ as a saw-tooth spin chain in detail \cite{Baniodeh}.  Motivated by some compounds such as $Bi_2Fe_4O_9$, F. C. Rodrigues {\it et al.} designed an interesting spin model for one stripe of the Cairo pentagonal Ising-Heisenberg lattice, then they investigated in detail zero-temperature phase transition for such model in Ref. \cite{Rodrigues}.  
 
 Quantum phase transitions have been one of the most interesting topics of strongly correlated systems during the last decade. It is basically a phase transition at zero temperature where the quantum fluctuations play the dominant role \cite{Arian1,Feng2007,Saadatmand,Valverde,Ananikian2012,Abgaryan1,Strecka1,Arian2}. Further studies to investigate
these quantum spin models have provided precise outcomes for the ground-state phase transition in the presence of an external magnetic field,  which can be induced through the exchange couplings  \cite{Sahoo1,Sahoo2,Giri,Hovhannisyan}. It is quite noteworthy that the ground state and thermodynamics of the spin ladders constituted by higher spins have been particularly examined as well \cite{Ivanov,Giri}.

Magnetization curves of low-dimensional quantum ferromagnets/antiferromagnets are topical issue of current research interest, in order to they often exhibit intriguing features such as magnetization plateaus. The spin-1/2 quantum chain in a transverse magnetic field \cite{Hida,Verkholyak}, the spin-1/2 quantum spin ladder \cite{Arian1,Cabra}, spin-1/2 Ising-Heisenberg diamond chain  in a transverse magnetic field \cite{Gu,Rojas2,RojasM,Ananikian2012,Abgaryan1,Strecka1} are a few exactly solved quantum spin models for which magnetization varies smoothly with rising absolute magnetic field until reaches its saturation magnetization. For the small quantum spin clusters, V. Ohanyan {\it et al.} investigated general non-commutativity features of the magnetization operator and Hamiltonian \cite{Ohanyan2015}. 
The specific heat of magnetic materials has attracted much attention over the past two decades, since it usually exhibits an anomalous thermal behavior by altering other parameters of the Hamiltonian such as coupling constants, spin exchange anisotropy and magnetic field etc. Such a parameter can be typically approximated under a certain thermodynamic condition by the Schottky theory \cite{Strecka,Karlova2016}. The associated round maximum of the specific heat, the so-called Schottky peak, has been experimentally detected in various magnetic compounds \cite{Bernu,Misguich,Helton}.

In solid state physics and Material science, most of the theoretical treatments are based on numerical techniques. Hence, an analytical approach to describe the ground state, magnetic and thermodynamic properties of the quantum spin systems such as the magnetization and specific heat, is definitely required. A promising method is the transfer-matrix formalism which has widely been applied to a number of strongly correlated systems at zero-temperature, as well as, low-temperature for studying the ground- and low-lying state properties of spin models. In the present paper, we theoretically describe how the transfer matrix method can be used to calculate the thermodynamic properties of a new specie of spin-1/2 ladder with Ising-Heisenberg interactions. 

\begin{figure}
\begin{center}
\resizebox{0.5\textwidth}{!}{%
\includegraphics{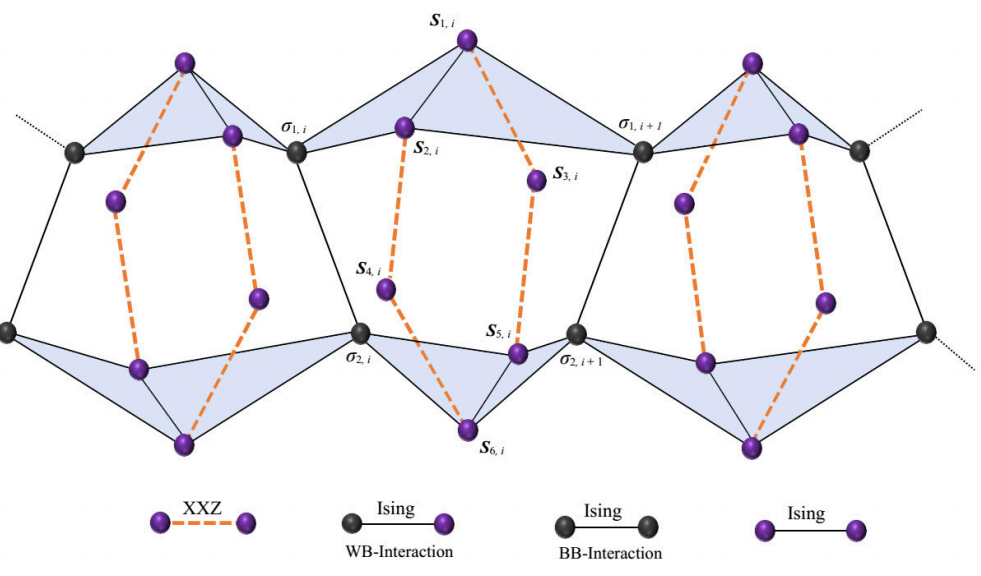}}
\caption{Schematic structure of a Ising-Heisenberg spin ladder of decanuclear spin-1/2 particle cages. WB and BB abbreviations denote Wing-Body  and Body-Body interactions, respectively.}
\label{fig:SpinLadder}
\end{center}
\end{figure}

In the present work, we are going to examine the magnetization and the specific heat for the interstitial half-spins of a spin ladder-type of 
decanuclear spin-1/2 particle cages in the presence of an external magnetic field at low temperature. The considered ladder with periodic boundary conditions is shown in Fig. \ref{fig:SpinLadder}. The great motivation to consider such a particular spin ladder with analytical Hamiltonian is to investigate theoretically some thermodynamic parameters like magnetization and specific heat of a so close spin model to real magnetic materials in terms of spin configuration that exhibits stimulating behaviors against magnetic field at low temperature.

The paper is organized as follows. In Sec. \ref{Model} we introduce the exactly solvable model. In Sec. \ref{TP}, we present the thermodynamic solution of the model with in the transfer-matrix formalism. In this section, we also numerically discuss the magnetization and specific heat of the model in the presence of an external homogeneous magnetic field. Some conclusions and future outlooks are briefly mentioned in Sec. \ref{conclusions}.
\section{Model and exact solution within the transfear matrix formalism}\label{Model}
The Hamiltonian of the spin model shown in Fig. \ref{fig:SpinLadder} can be expressed as
\begin{equation}
\begin{array}{lcl}
H= \sum\limits_{i=1}^N \Big[-\sum\limits_{j=1}^4{J}_H ({\bf S}_{j,i}\cdot{\bf S}_{j+2,i})_{\Delta}+J_{Is} \big(S_{1,i}^z S_{2,i}^z + S_{5,i}^z S_{6,i}^z\big)\\
J_{Is} \big[(\sigma_{1,i}^z+ \sigma_{1,i+1}^z) (S_{1,i}^z + S_{2,i}^z) + (\sigma_{2,i}^z+ \sigma_{2,i+1}^z) (S_{5,i}^z + S_{6,i}^z)+\\
\sigma_{1,i}^z \sigma_{2,i}^z + \sigma_{2,i}^z \sigma_{2,i+1}^z\big]
-g\mu_B B_z\big(\sum\limits_{j=1}^{6} S_{j,i}^z + \frac{1}{2}\sum\limits_{j=1}^{4} \sigma_{j,i}^z\big)\Big],
\end{array}
\end{equation}
where $N$ is the number of blocks and
\begin{equation}
\begin{array}{lcl}
 ({\bf S}_{j,i}\cdot{\bf S}_{j+2,i})_{\Delta}=S_{j,i}^xS_{j+2,i}^x+S_{j,i}^yS_{j+2,i}^y+\Delta S_{j,i}^zS_{j+2,i}^z,
\end{array}
\end{equation}
corresponds to the interstitial anisotropic Heisenberg spins coupling ${J}_H$ and $\Delta$, while the nodal spins localized on the 
$i$-th rung are representing by Ising-type exchanges $J_{Is}$.
 $2S^\alpha=\sigma^\alpha$ for which ${\sigma^{\alpha}}=\lbrace {\sigma}^x, {\sigma}^y, {\sigma}^z \rbrace$ are Pauli operators (with $\hbar=1$). ${B}_z$ is applied homogeneous magnetic field in the $z$-direction. The gyromagnetic ratio is taken to be $g=2.42$ ($\mu_B=1$) in the plots drawn in this paper. 
 
The cornerstone of our further calculations is based on the commutation relation between different block Hamiltonians $[h_i, h_j] = 0$, which will allow us to characterize the partition function of the ladder under consideration and represent it as a product over block partition functions
%\begin{equation}\label{PF}
${Z}=Tr\Big[\displaystyle\prod_{i=1}^{N}\exp(-\beta h_{i})\Big],$
%\end{equation}
where $\beta=\frac{1}{k_{B}T}$, $k_{B}$ is the Boltzmann’s constant and $T$ is the temperature.
 In the two qubit standard eigenbasis of the composite spin operators $\{ \sigma_{1,i}^{z},\sigma_{2,i}^{z},\sigma_{1,i+1}^{z},\sigma_{2,i+1}^{z}\} $ of the two consecutive rungs of the block $i$, we can consider the following matrix representation to  formulate partition function
  ${Z}$ as
\begin{equation}\label{PF}
\begin{array}{lcl}
{Z}=Tr\big[ \langle \sigma_{1,1}^{z}\sigma_{2,1}^{z}\vert\mathcal{T}\vert \sigma_{1,2}^{z}\sigma_{2,2}^{z}\rangle\cdots\langle \sigma_{1,N}^{z}\sigma_{1,N}^{z}\vert \mathcal{T}\vert \sigma_{2,N+1}^{z}\sigma_{2,N+1}^{z}\rangle \big],
\end{array}
\end{equation}
where $\sigma_{j,i}^{z}=\pm1$, and under the periodic boundary conditions we have $\sigma_{j,N+1}=\sigma_{j,1}$
. We can figure out the $4\times4$ transfer matrix $\mathcal{T}$ as follows
\begin{equation}\label{TrM}
\begin{array}{lcl}
\mathcal{T}(i)=\langle \sigma_{1,i}^{z}\sigma_{2,i}^{z}\vert\exp(-\beta h_{i})\vert \sigma_{1,i+1}^{z}\sigma_{2i,+1}^{z}\rangle=\\
\sum\limits_{k=1}^{64}\exp\big[-\beta\mathcal{E}_k(\sigma_{1,i}^{z}\sigma_{2,i}^{z},\sigma_{1,i+1}^{z}\sigma_{2,i+1}^{z})\big].
\end{array}
\end{equation}
$\mathcal{E}_k$ denotes eigenvalues of the unit block Hamiltonian. Since we seek eigenvalues of the transfer matrix in the thermodynamic limit $N\rightarrow\infty$, the largest eigenvalue $\Lambda_{max}$ has the most effect on the thermodynamic properties of the system, whereas other three smaller eigenvalues are almost effectless and their contribution can be completely neglected.
Hence, the free energy per block can be obtained from the largest eigenvalue of the transfer matrix (\ref{TrM}) as
 \begin{equation}\label{FreeE}
 \begin{array}{lcl}
 f=-\frac{1}{\beta}\lim\limits_{N\rightarrow \infty}\ln\frac{1}{N}{Z}=-\frac{1}{\beta}\ln\Lambda_{max}.
 \end{array}
 \end{equation}
 
 \begin{figure}
\begin{center}
\resizebox{0.4\textwidth}{!}{%
\includegraphics{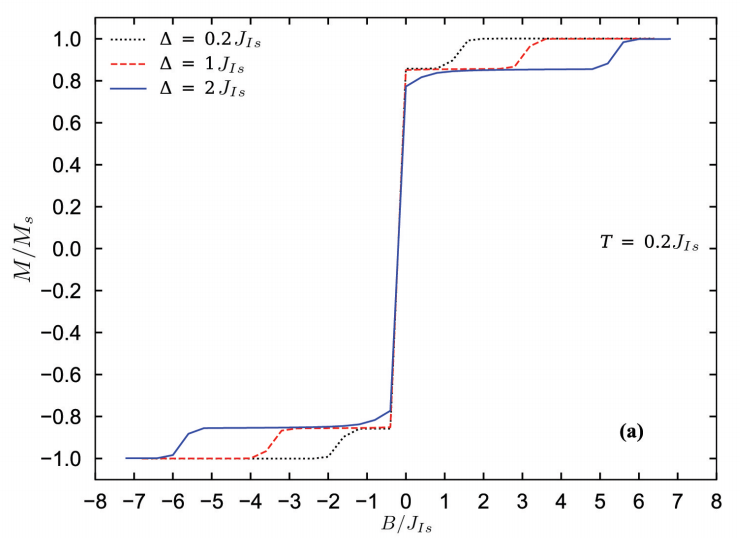}
}
\resizebox{0.4\textwidth}{!}{%
\includegraphics{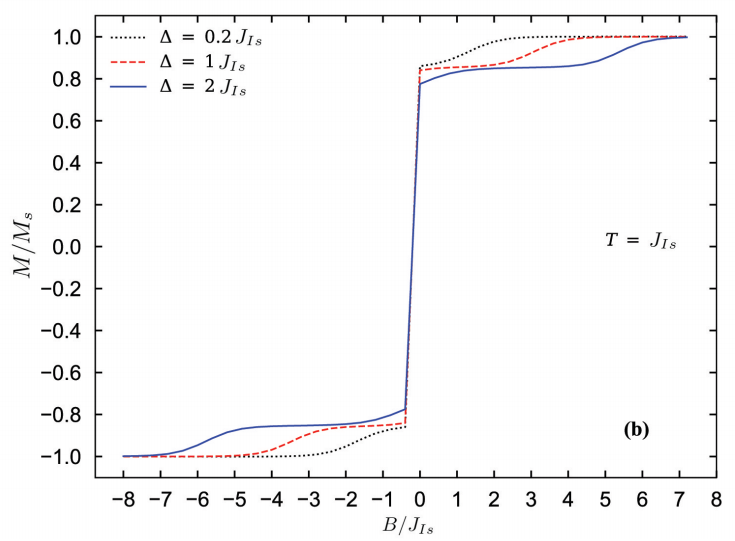}
}
\resizebox{0.4\textwidth}{!}{%
\includegraphics{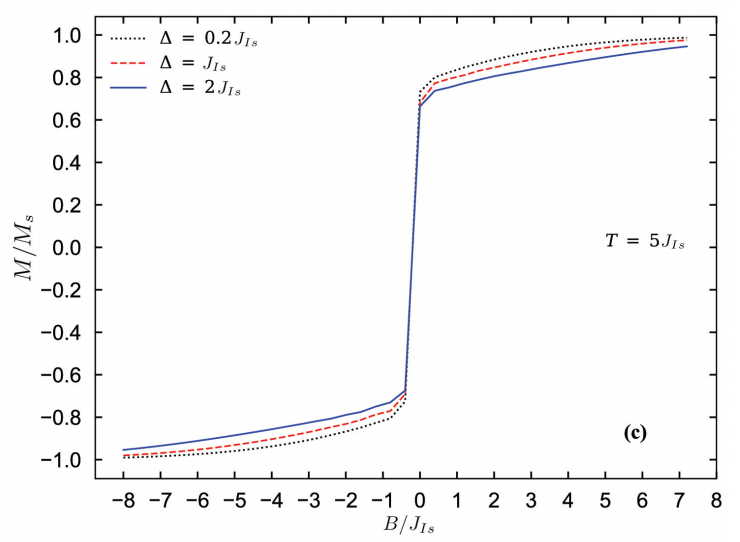}
}
\caption{Magnetization per saturation value $M/M_s$ as a function of external magnetic field $B/J_{Is}$ for fixed value of $J_{H}=1.5J_{Is}$ and three different coupling constant $\Delta=0.2J_{Is}$, $\Delta=J_{Is}$  and $\Delta=2J_{Is}$, at (a) low temperature  $T= 0.2J_{Is}$, (b) $T=J_{Is}$, and (c) $T=5J_{Is}$.}
\label{fig:Mat}
\end{center}
\end{figure}
 \section{Results and discussion}\label{TP}
Now one can utilize the thermodynamic relations to evaluate various quantities that would be investigated. Specific heat, entropy, and magnetization per block can be consequently defined as
\begin{equation}\label{TParameters}
\begin{array}{lcl}
{M}=-\Big(\frac{\partial f}{\partial B}\Big)_{T}, \quad {S}=-\Big(\frac{\partial f}{\partial T}\Big)_{B}, \quad {C}=-T\Big(\frac{\partial^2 f}{\partial T^2}\Big)_{B}.
\end{array}
\end{equation}
Figure \ref{fig:Mat} illustrates the magnetic field dependences of magnetization in the unit of its saturation. Figure \ref{fig:Mat} (a) demonstrates the magnetization per block against the magnetic field at low temperature, and Figs. \ref{fig:Mat} (b) and \ref{fig:Mat} (c) display this quantity versus the magnetic field at higher temperatures. It is quite obvious that the magnetization curve shows an intermediate plateau at $\frac{5}{6}$ of saturation magnetization . The plateau becomes wider upon increasing the coupling constant $\Delta/J_{Is}$, namely, for higher values of 
$\Delta/J_{Is}$, magnetization jumps from the $\frac{5}{6}-$plateau to the saturation magnetization at higher absolute values of the magnetic field. Interestingly, as the temperature increases the plateau gradually disappears until at high temperature the magnetization behaves as a smooth  logarithmic curve without plateau. It its also noteworthy that with increase of the temperature the magnetization curves coincide together, in this situation the spin exchange anisotropy $\Delta/J_{Is}$ acts as a resistance factor of this coincidence.
 \begin{figure}
\begin{center}
\resizebox{0.5\textwidth}{!}{%
\includegraphics{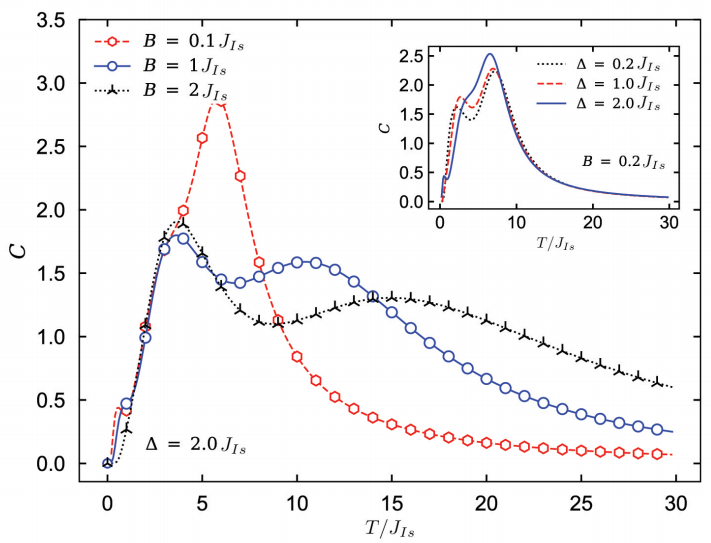}
}
\caption{The specific heat of the introduced spin ladder as a function of the temperature for various fixed values of the magnetic field 
$B=0.1J_{Is}$, $B=J_{Is}$ and $B=2J_{Is}$, where other parameters are taken as  $\Delta=2J_{Is}$ and $J_{H}=1.5J_{Is}$. The inset displays the temperature dependence of specific heat in the presence of a weak magnetic field for three different dimensionless coupling constant $\Delta=0.2J_{Is}$, $\Delta=J_{Is}$  and $\Delta=2J_{Is}$.}
\label{fig:SHeat}
\end{center}
\end{figure}
Now let us examine the effects of anisotropy $\Delta/J_{Is}$ and the magnetic field on the temperature dependence of the specific heat. To this end, we display in Fig. \ref{fig:SHeat} the temperature dependence of the specific heat for the model under consideration for the several fixed values of the magnetic field. When the temperature growths from zero the specific heat increases till a small peak arises ($T<J_{Is}$) in weak magnetic field (red curve). By inspecting Figs. \ref{fig:Mat} and \ref{fig:SHeat} and a precise comparing, one can realize that this peak arises when the magnetization plateau appears. It could be expected that for very weak magnetic fields ($B\ll J_{Is}$) and/of high anisotropies ($\Delta>2J_{Is}$, for more detail see the inset), the specific heat curve manifests a Schottky-type maximum that strongly depends on the strength of the magnetic field.
With further increase of the magnetic field, this thermodynamic quantity exhibits more peaks. Actually, in the presence of the higher magnetic fields, the specific heat of the model displays a small peak at finite low temperatures. When the temperature increases, a steep increase appears where second peak is created, and in turn third peak arises that is highest peak. 
The latter peak diminishes upon increasing the magnetic field. With further increase of the magnetic field the height of such a peak gradually decreases until becomes smaller than second peak and moves toward higher temperatures. This phenomenon is in accordance with the ferromagnetic-antiferromagnetic quantum phase transition. As an outstanding result, here the specific heat shows three separated peaks that are strongly dependent on the magnetic field. Besides, first and second peaks undergo considerable changes under magnetic field variations, as a matter of fact, they merge together upon increasing magnetic field. These changes in specific heat behavior are compatible with the magnetic  alternations of the system.

The inset of  Fig. \ref{fig:SHeat} shows the temperature dependence of the specific heat for various fixed values of the coupling constant $\Delta/J_{Is}$ in the presence of a weak magnetic field ($B=0.2J_{Is}$). Obviously, the specific heat curves coincide at high temperature when a weak magnetic field is applied. At low temperature, this quantity depicts different behavior with respect to the coupling constant $\Delta/J_{Is}$ changes. To clarify this point, one can see that when $\Delta/J_{Is}$ decreases the smallest peak of specific heat curve gradually disappears, whereas the specific heat resists to keep its larger peaks. This adventure indicates that the system undergoes one another quantum phase transition that may occur under this specific thermodynamic condition. At high temperature the specific heat decreases and gradually goes to zero. 

\section{Conclusions}\label{conclusions}
In this paper, we have theoretically investigated thermodynamic properties of a new spin ladder-type consisting of spin-1/2 particle cages for which unit blocks include four pivalate ligands bridge wing-body spin centers within the butterflies.  We consider two extra interstitial half-spins in each cage which have XXZ Heisenberg interaction with spins localized on the bodies of butterflies. To do so, we have examined the magnetization process and specific heat of this model by means of solution within the transfer-matrix formalism. 

 In terms of numerical investigations, we understood that the magnetization has a plateau at $\frac{5}{6}$ of the saturation magnetization. This plateau rigorously depends on the temperature and spin exchange anisotropy. Furthermore, it has been demonstrated that in the presence of the weak magnetic fields, the specific heat curve manifests three separated peaks temperature dependence. Magnetic field and the anisotropy variations can remarkably alter the shape and the temperature position of these peaks.
 Namely, the magnetic field increment (the anisotropy decrement) leads to vanishing the smallest peak, hence, the specific heat has just a double-peak. The thermal excitation of low-lying energy causes this anomalous double-peak. Thus, the specific heat variations with respect to the temperature are in a good accordance with the low-temperature magnetization response to the anisotropy and temperature changes, this may indicates the ground-state phase transition related to the magnetization jump.
 
 Further applications of the method discussed in this paper pave the way to set and interpret the numerical and analytical expressions for
utilizing transfer matrix formalism in realistic scenarios and comparison of the results with the experimental data obtained from investigating novel magnetic materials with the similar spin configuration, which can play an important role to better understand features of more complicated magnetic materials and their thermodynamic properties.

\section*{Acknowledgments}
H. Arian Zad acknowledges the receipt of the grant from the ICTP Affiliated Center Program AF-04.

%\section*{References}

\bibliography{}

\begin{thebibliography}
\bibliography{}

 \bibitem{Kitaev}
 G. Vidal,  J. I. Latorre,  E. Rico and  A. Kitaev, \href{https://dx.doi.org/10.1103/PhysRevLett.90.227902}{ {\it Phys. Rev. Lett.}  \textbf{90}, 227902 (2003).}

\bibitem{Gu}
  B. Gu and  G. Su,  \href{https://dx.doi.org/10.1103/PhysRevB.75.174437}{{\it Phys. Rev. B}  \textbf{75}, 174437 (2007).}

\bibitem{Dillenschneider}
  R. Dillenschneider,   \href{https://dx.doi.org/10.1103/PhysRevB.78.224413}{{\it Phys. Rev. B} \textbf{78}, 224413 (2008).}

\bibitem{Ivanov}
  N. B. Ivanov, J. Richter and  J. Schulenburg, \href{https://dx.doi.org/10.1103/PhysRevB.79.104412}{{\it Phys. Rev. B} \textbf {79}, 104412 (2009).}

\bibitem{Werlang}
  T. Werlang,  C. Trippe,  G. A. P. Ribeiro and  G. Rigolin,  \href{https://dx.doi.org/10.1103/PhysRevLett.105.095702}{ {\it Phys. Rev. Lett.}  \textbf {105}, 095702 (2010).}
  
\bibitem{Sachdev}
  S. Sachdev,  {\it Quantum Phase Transitions} (Cambridge University Press, Cambridge, (2011)).

\bibitem{Rojas2}
 O. Rojas,  M. Rojas,  N. S. Ananikian and  S. M. D. Souza, \href{https://dx.doi.org/10.1103/PhysRevA.86.042330}{ {\it Phys. Rev. A}  \textbf{86}, 042330 (2012).}

\bibitem{RojasM}
  J. Torrico,  M. Rojas,  S. M. D. Souza,  O.  Rojas and  N. S. Ananikian, \href{https://dx.doi.org/10.1209/0295-5075/108/50007}{{\it Europhys. Lett.} \textbf{108}, 50007 ( 2014).}

\bibitem{Strecka}
  J. Strecka,  R. C. Alecio,  M. Lyra and  O. Rojas,  \href{https://dx.doi.org/10.1016/j.jmmm.2016.02.095} {{\it J. Magn. Magn. Mater.} \textbf{409}, 124 (2016).}

\bibitem{Koga}
  A. Koga,  S. Kumada,  N. Kawakami and  T. Fukui,  \href{https://dx.doi.org/10.1143/JPSJ.67.622}{ {\it J. Phys. Soc. Jpn.} \textbf{67}, 622 (1998).}

\bibitem{Okamoto}
  K. Okamoto,  N. Okazaki and  T. Sakai,  \href{https://dx.doi.org/10.1143/JPSJS.71S.196} { {\it J. Phys. Soc. Jpn.} \textbf{71}, 196 (2002).}

\bibitem{Muller}
 M. Muller,  T. Vekua and  H. J. Mikeska  \href{https://dx.doi.org/10.1103/PhysRevB.66.134423}{{\it Phys. Rev. B} \textbf {66}, 134423 (2002).} 

\bibitem{Vuletic}
 T. Vuletic,  B. K. Hamzic,  T. Ivek, S. Tomic,  B. Gorshunov,  M. Dressel and  J. Akimitsu,  \href{https://doi.org/10.1016/j.physrep.2006.01.005}{ {\it Phys. Rep.} \textbf {428}, 169 ( 2006).}

\bibitem{Notbohm}
 S. Notbohm, {\it Spin Dynamics of Quantum Spin-Ladders and Chains}, PhD thesis, University of St Andrews,  (2007).

\bibitem{Buttner}
 S. Chen,  H. Buttner and  J. Voit,   \href{https://doi.org/10.1103/PhysRevB.67.054412}{{ Phys. Rev. B} \textbf{67}, 054412  (2003).}

\bibitem{Blundell}
 S. A. Blundell and  M. D. N. Regueir, \href{https://doi.org/10.1140/epjb/e2003-00054-2}{ { Eur. Phys. J. B} \textbf{31},  453 (2003).}

\bibitem{Bacq}
 O. L. Bacq,   A. Pasturel,  C. Lacroix and  M. D. N. Regueiro, \href{https://doi.org/10.1103/PhysRevB.71.014432}{{ Phys. Rev. B} \textbf{71}, 014432 (2005).}
 
 \bibitem{Arian1}
 H. Arian Zad and  N. Ananikian,  \href{https://doi.org/10.1088/1361-648X/aa8dd0}{{ J. Phys. Condens. Matt.} \textbf{29}, 455402 (2017).}

\bibitem{Feng2007}
X. Y.  Feng, G. M. Zhang and T. Xiang, \href{https://doi.org/10.1103/PhysRevLett.98.087204}{ {\it Phys. Rev. Lett.} {\bf 98}, 087204 (2007).}

\bibitem{Drillon1}
M. Drillon, M. Belaiche, P. Legoll, J. Aride, A. Boukhari and  A. Moqine, \href{https://doi.org/10.1016/0304-8853(93)90860-5}{ {\it J. Magn. Magn. Mater.} {\bf 128}, 83 (1993).}
 
 \bibitem{Okamoto1999}
K. Okamoto, T. Tonegawa and M. Kaburagi and M. Kaburagi, \href{https://doi.org/10.1088/0953-8984/11/50/336}{{\it J. Phys.: Condens. Matter.} {\bf 11}, 10485 (1999).}

\bibitem{Okamoto2003}
K. Okamoto, T. Tonegawa and M. Kaburagi, \href{https://doi.org/10.1088/0953-8984/15/35/307}{{\it J. Phys.: Condens. Matter.} {\bf 15}, 5979 (2003).}

\bibitem{Uematsu}
 D. Uematsu and M. Sato, \href{https://doi.org/10.1143/JPSJ.76.084712}{{\it J. Phys. Soc. Jpn.} {\bf 76}, 084712 (2007).}

\bibitem{Kikuchi1}
H. Kikuchi, Y. Fujii, M. Chiba, S. Mitsudo, T. Idehara and T.Kuwai, \href{https://doi.org/10.1016/j.jmmm.2003.12.619}{{\it J. Magn. Magn. Mater.} {\bf 272}, 900  (2004)}.

\bibitem{Kikuchi2}
H. Kikuchi, Y. Fujii, M. Chiba, S. Mitsudo, T. Idehara, T. Tonegawa, K. Okamoto, T. Sakai, T. Kuwai and H. Ohta
\href{https://doi.org/10.1103/PhysRevLett.94.227201}{{\it Phys. Rev. Lett.} {\bf 94},  227201 (2005).}

\bibitem{Baniodeh}
A. Baniodeh, N. Magnani, Y. Lan, G. Buth, C. E. Anson, J. Richter, M. Affronte, J. Schnack and A. K. Powell, 
\href{https://doi.org/10.1038/s41535-018-0082-7}{ {\it npj Quant. Mater} {\bf 3}, 10 (2018).} 

\bibitem{Rodrigues}
 F. C. Rodrigues, S. M. de Souza and O. Rojas, \href{https://doi.org/10.1016/j.aop.2017.02.005} {{\it Annals of Phys.} {\bf 379}, 1 (2017).}


\bibitem{Saadatmand}
S. N. Saadatmand, B. J. Powell and I. P. McCulloch, \href{https://doi.org/10.1103/PhysRevB.91.245119}{{\it Phys. Rev. B} {\bf 91}, 245119 (2015).}

\bibitem{Valverde}
J.S. Valverde, O. Rojas and S. M. de Souza,  \href{https://doi.org/10.1088/0953-8984/20/34/345208}{{\it J. Phys.: Condens. Matt.} {\bf 20}, 345208 (2008).}

\bibitem{Ananikian2012}
N. S. Ananikian, L. N. Ananikyan, L. A. Chakhmakhchyan and O. Rojas, {J. Phys.: Condens. Matt.} \href{https://dx.doi.org/10.1088/0953-8984/24/25/256001}{\textbf{24},  256001(2012).}

\bibitem{Abgaryan1}
 V. S. Abgaryan, N. S. Ananikian, L. N. Ananikyan and V. Hovhannisyan, {Solid State Comm.} \href{https://dx.doi.org/10.1016/j.ssc.2014.11.013}{ \textbf{203},  5 (2015)};
V. S. Abgaryan, N. S. Ananikian, L. N. Ananikyan and V. Hovhannisyan, {Solid State Comm.} \href{https://dx.doi.org/10.1016/j.ssc.2015.10.003}{ \textbf{224}, 15 (2015) .}

\bibitem{Strecka1}
  O. Rojas,  M. Rojas,  S. M. D. Souza,  J. Torrico,  J. Strecka and  M. L. Lyra, \href{https://dx.doi.org/10.1016/j.jmmm.2016.02.095}{{ Physica  A} {\bf 486},  367 (2017).}
 
 \bibitem{Arian2}
 H. Arian Zad and  N. Ananikian,  \href{https://doi.org/10.1088/1361-648X/aab644}{{\it J. Phys. Condens. Matt.} {\bf 30}, 165403 (2018).}

\bibitem{Sahoo1}
V. M. L. D. Prasad Goli, S. Sahoo, S. Ramasesha and D.Sen,  \href{http://dx.doi.org/10.1088/0953-8984/25/12/125603}{{\it J. Phys.: Condens. Matt.} {\bf 25}, 125603 (2013).}
 
 \bibitem{Sahoo2}
S. Sahoo, V. M. L. D. Prasad Goli, D.Sen and S. Ramasesha, \href{http://dx.doi.org/10.1088/0953-8984/26/27/276002}{\it J. Phys.: Condens. Matt.} {\bf 26}, 276002 (2014).
 
  \bibitem{Giri}
G. Giri, D. Dey, M. Kumar, S. Ramasesha, and Z. G. Soos,  \href{https://doi.org/10.1103/PhysRevB.95.224408}{ {\it Phys. Rev. B} {\bf 95}, 224408 (2017).}
 
 \bibitem{Hovhannisyan}
 V. Hovhannisyan, J. Strecka and N. Ananikian, \href{https://doi.org/10.1088/0953-8984/28/8/085401 }{{\it J. Phys.: Condens. Matt.} {\bf 28}, 085401(2016).}
 
\bibitem{Hida}
K. Hida,  \href{https://doi.org/10.1143/JPSJ.63.2514}{{\it J. Phys. Soc. Jpn.} \textbf {63}, 2514 (1994).}

 \bibitem{Verkholyak}
T. Verkholyak, J. Strecka, M. Jascur and J. Richter, \href{https://doi.org/10.1140/epjb/e2011-10681-5}{{\it Eur. Phys. J. B} {\bf 80}, 433 (2011).}

\bibitem{Cabra}
D. C. Cabra, A. Honecker and P. Pujol, \href{https://doi.org/10.1103/PhysRevLett.79.5126}{{\it Phys. Rev. Lett.} \textbf{79}, 5126 (1997).}
 
\bibitem{Ohanyan2015}
V. Ohanyan, O. Rojas, J. Strecka and S. Bellucci, \href{https://doi.org/10.1103/PhysRevB.92.214423}{{\it Phys. Rev. B}{ \bf 92}, 214423 (2015).}

\bibitem{Karlova2016}
 K. Karlova, J. Strecka and T. Madaras,  \href{https://doi.org/10.1016/j.physb.2016.01.033}{{\it Physica B} {\bf 488}, 49 (2016).}

\bibitem{Bernu}
 G. Misguich and B. Bernu, \href{https://doi.org/10.1103/PhysRevB.71.014417}{{\it Phys. Rev. B} {\bf 71}, 014417 (2005).}

\bibitem{Misguich}
  G. Misguich and P. Sindzingre, \href{https://doi.org/10.1140/epjb/e2007-00301-6}{{\it Eur. Phys. J. B}{\bf 59}, 305 (2007).}

\bibitem{Helton}
J. S. Helton, K. Matan, M. P. Shores, E. A. Nytko, B. M. Bartlett, Y. Yoshida, Y. Takano, A. Suslov, Y. Qiu, J.-H. Chung, D. G. Nocera and Y. S. Lee, \href{https://doi.org/10.1103/PhysRevLett.98.107204}{{\it Phys. Rev. Lett.}{\bf 98}, 107204 (2007).}
 
\bibitem{Sheikh}
J. A. Sheikh, A. Adhikary, H. S. Jena, S. Biswas, and S. Konar, \href{https://doi.org/10.1021/ic402673v}{\it Inorg. Chem.} {{\bf 53}, 1606 (2014).}

\end{thebibliography}

\end{document}